# Primordial Black Holes as Heat Sources for Living Systems with Longest Possible Lifetimes

*C Sivaram*

Indian Institute of Astrophysics, Bangalore, 560 034, India

Telephone: +91-80-2553 0672; Fax: +91-80-2553 4043

e-mail: sivaram@iiap.res.in

*Kenath Arun*

Christ Junior College, Bangalore, 560 029, India

Telephone: +91-80-4012 9292; Fax: +91-80- 4012 9222

e-mail: kenath.arun@cjc.christcollege.edu

*Kiren O V*

Christ Junior College, Bangalore, 560 029, India

Telephone: +91-80-4012 9292; Fax: +91-80- 4012 9222

e-mail: kiren.ov@cjc.christcollege.edu

**Abstract:** Just forty years ago, Hawking wrote his famous paper on primordial black holes (PBH). There have been since innumerable discussions on the consequences of the existence of such exotic objects and ramifications of their properties. Here we suggest that PBH's in an ever expanding universe (as implied by dark energy domination, especially of a cosmological constant) could be the ultimate repository for long lived living systems. PBH's having solar surface temperatures would last $10^{32}$ years as a steady power source and should be considered in any discussion on exobiological life.

Forty years ago, in March 1974, Hawking published his seminal paper titled 'Black Hole Explosions' (Hawking, 1974). He pointed out that although classically particles cannot escape from a black hole (they are all trapped in the strong gravitational potential well of the event horizon) in quantum theory there is a small probability (like a quantum tunnelling process) for particles to be found outside the event horizon.

This is similar to the decay of heavy nuclei, wherein alpha particles with too little energy to escape the nuclear potential barrier, can nevertheless tunnel their way out of the nucleus (the quantum probability depending on the particle energy and the charge of the decaying nuclei). So the radioactive nucleus slowly decays. So particles can quantum mechanically tunnel their way out of the black hole and moreover it turns out that the energy spectrum of the emitted particles corresponds to a black body (Hawking 1975; Page 1976) with temperature depending inversely on the mass. Thus we have:

$$T_{BH} = \frac{\hbar c^3}{8\pi G M_{BH}} \qquad \ldots (1)$$

$T_{BH}$, $M_{BH}$ are the black hole horizon temperature and mass of black hole respectively. For a solar mass black hole this temperature (at the horizon) is hardly a micro-kelvin.

The Hawking flux (which is essentially a black body radiation from the horizon at temperature $T_{BH}$) is given as:

$$f_{BH} = \sigma A_H T_{BH}^4 \qquad \ldots (2)$$

$A_H$ being the horizon surface area, which is $4\pi \left( \frac{2GM_{BH}}{c^2} \right)^2$, $\sigma$ is the Stephan-Boltzmann constant.

The Hawking flux thus depends on $M_{BH}$ as:

$$f_{BH} \sim \frac{\hbar c^6}{G^2 M_{BH}^2} \qquad \ldots (3)$$

For a solar mass black hole, this Hawking flux is only a yocto-watt ($10^{-24}$W). Thus the evaporation time (life-time of black holes to decay by Hawking radiation) of the isolated black hole is given as:

$$t_{BH} = \frac{G^2 M_{BH}^3}{\hbar c^4} \qquad \ldots (4)$$

So that a solar mass black hole would ultimately evaporate on a time scale of $10^{64}$ years! This implies that Hawking radiation is completely negligible and unobservable for all black holes currently forming astrophysically in the universe. (Sivaram, 2000)

However Hawking pointed out that in the very early universe with conditions of extremely high density and temperature, primordial black holes (PBH's) could form of any mass (this mass depending on epoch (in the early universe) when they formed.)

This mass is given as (at an epoch t, in the early universe) (Sivaram, 2001; 1983):

$$M_{BH} = \frac{c^3 t}{8\pi G} \qquad \ldots (5)$$

It is thus of great interest that PBH's of a mass around $10^{15}$g (a billion tonnes) would have evaporation time scales of ten billion years, they would just now be releasing their energy explosively in the form of gamma rays (their horizon temperature corresponding to a trillion degrees) with an energy release of around $10^{30}$ Joules.

It is convenient to express equations (1)-(4), in terms of the corresponding values for a Planck mass black hole, the Planck mass being the quantum black hole with the lowest mass of

$$M_{pl} = \left(\frac{\hbar c}{G}\right)^{1/2} \sim 2 \times 10^{-5} g \qquad \ldots (6)$$

Thus:

$$T_{BH} = 10^{32}\left(\frac{M_{pl}}{M_{BH}}\right)K \qquad \ldots (7)$$

$$f_{BH} = 10^{52} W \left(\frac{M_{pl}}{M_{BH}}\right)^2 \qquad \ldots (8)$$

$$t_{BH} = 10^{-43} \sec \left( \frac{M_{BH}}{M_{Pl}} \right)^3 \quad \ldots (9)$$

Thus a PBH, with a billion ton mass, with a lifetime of Hubble age coincidentally emits a power of $10^{13}$W, just what is consumed by mankind! So if we could find such a PBH, all our power problems would be solved for a billion years!

The above equations imply that there could be PBH's with solar temperatures, emitting most of their energy in the optical or IR. Such Hawking black holes would have a life time of the order of $10^{32}$ years (equations (7) and (9)), but of course their luminosity would be of the order of milli-watts. They would be one micron across, but material orbiting them even 0.1 cm away would receive a flux comparable to solar flux on earth. So if there are suitable molecules in the materials orbiting such a PBH it could host at least bacterial life and other microorganisms (in the micron size range). At larger distances from the PBH, flux would be smaller. At one meter distance it would be only a few micro-watts. But the important or unusual aspect is that such life (whatever be its nature) can have a continuous power source for $10^{32}$ years!

In an ever expanding universe, this would be the ultimate repository of life. The time scales are much longer than those involving red dwarfs, in fact a pentillion times longer!

It appears that life need not be confined only in the neighbourhood of massive stars within a narrow mass range of 0.8 – 1.3 solar masses and that too within their habitable zones (where water can exist as a liquid).

Indeed the first system of 'planets' was observed around a neutron star. Also recently it has been advocated that white dwarfs in their cooling phase, which can last several billion years (with surface temperatures hovering near solar surface temperatures for this period) can host life on planets orbiting them at close distances. Such planets would have a heat flux comparable to what earth gets from the sun and even could have 'blue skies'.

Again the longest lasting forms of life could perhaps be found around red dwarfs, low mass stars (0.2 – 0.1, solar mass) which could have life times of ten trillion years, the so called stellar Methuselah! These stars would not become red giants also, but merely brighten up to about the Sun's luminosity after a trillion years!

So the oldest forms of life or 'oldest' fossils for that matter (if any) are likely to be present (if at all!) around old red dwarfs. In an ever expanding universe (like the one we are now supposed to inhabit and supposedly driven by the negative pressure of dark energy leading to faster and faster expansion), life around red dwarfs could still be present even after fifty trillion years! (Sivaram and Sastry 2005)

On earth bacterial life was the earliest, billions of years ago and is still dominant. It has out lived dinosaurs and other behemoths and would in all likelihood outlive humankind (our species still being most vulnerable to new bacterial and viral strains). So it looks singularly appropriate that in future, the longest living life forms would again be micron or nano sized organisms, orbiting PBH's and lasting well beyond $10^{32}$ years! (Sivaram et al, 2014)